\documentclass[runningheads]{llncs}
\usepackage{graphicx}
\pdfoutput=1

\begin{document}
\title{SonoUno web: an innovative user centred web interface\thanks{Supported by the Project REINFORCE (GA 872859) with the support of the EC Research Innovation Action under the H2020 Programme SwafS-2019-1the REINFORCE (\url{www.reinforceeu.eu})}}
%
%
\author{Gonzalo De La Vega \and
Leonardo Martin Exequiel Dominguez \and
Johanna Casado\inst{1}\inst{2}\orcidID{0000-0001-9528-5034} \and Beatriz Garc\'{i}a\inst{1}\inst{3}}
\authorrunning{G. De La Vega et al.}
%
\institute{Instituto en Tecnolog\'{i}as de Detecci\'{o}n y Astropart\'{i}culas (CNEA,CONICET, UNSAM), Mendoza, Argentina \\
\email{johanna.casado@um.edu.ar}\\
\url{http://www.sonouno.org.ar} \and
Instituto de Bioingenier\'{i}a, Facultad de Ingenier\'{i}a, Universidad de Mendoza, Argentina \and
Universidad Tecnol\'{o}gica Nacional, Argentina}
\maketitle              
\begin{abstract}
Sonification as a complement of visualization is been under research for decades as a new ways of data deployment. ICAD conferences, gather together specialists from different disciplines to discuss about sonification. Different tools as sonoUno, starSound and Web Sandbox are attempt to reach a tool to open astronomical data sets and sonify it in conjunction to visualization. In this contribution, the sonoUno web version is presented, this version allows user to explore data sets without any installation. The data can be uploaded or a pre-loaded file can be opened, the sonification and the visual characteristics of the plot can be customized on the same window. The plot, sound and marks can be saved. The web interface were tested with the main used screen readers in order to confirm their good performance.

\end{abstract}
\section{Introduction}

The need to explore data sets beyond the visual field has led the community to study new ways to represent it, this is the case of sonification. In this sense, since 1992 the ICAD conferences\cite{icadWeb} has existed bringing together scientists from different fields to discuss sonification, how people perceive it and how it can be used. Related to sonification Phillips and Cabrera\cite{sonifworkstation} present a sonification workstation; and related to astronomy Shafer et.al.\cite{solarharmonics} and Garc\'{i}a Riber\cite{sonifigrapher} develop specific projects to sonify solar harmonics and light curves.

During the past years some sonification programs were created as tools to make possible the multimodal exploration of visual and audio graphs, this is the case of xSonify\cite{wandaetal2011}, Sonification Sandbox\cite{sandbox2007}, Sonipy\cite{sonipy-proc,sonipy-web}, StarSound\cite{starsound} and SonoUno\cite{sonouno-github}. All are standard alone software that requires you to download a package and install it. Related to the possibility of analyze data with sonification, Díaz-Merced \cite{wanda2013} in her thesis, using the standard alone sonification software xSonify, concluded that sonification as a complement to visual display augments the detection of features in the data sets under analysis.

Given the complexity to use the available standard alone software and to avoid errors and problems during the software installation, the idea of a sonification software working through the web began to make sense. TwoTone\cite{twotone}, TimeWorkers\cite{chafe}, Sonification Blocks\cite{sonifblocks} and Web Sandbox\cite{sandbox-web} are different attempts to make it real, but none of them allow the end user to explore, make choices about the configuration and how they want to display the data and functionalities. In this sense, we present in this contribution a graphic user interface available in the web that presents the same user centred framework and almost same functionalities of sonoUno \cite{casado2019iau} desktop software.

The sonoUno software, in its web and desktop versions, is a public tool to display, sonify and apply mathematical functions to any data set. The actual original application of the software is to astronomical data, but it can be used with any type of data presented in two or more columns (csv or txt) files. SonoUno presents a user centred approach from the beginning, first with a theoretical framework, second with  focus group sessions and then with a community of people that kindly test the software and send the feedback to the developers\cite{casado2020iau}.

The sonoUno web interface was tested in different operative system and with different screen readers. This work was partially financiated by the Project REINFORCE (GA 872859) with the support of the EC Research Innovation Action under the H2020 Programme SwafS-2019-1 the REINFORCE \url{www.reinforceeu.eu}.

\section{Methodology}

Taking in mind that the end user must have the ability to choose, configure and decide how they want to explore their data sets, this project requires the use of HTML, JavaScript, CSS and ARIA (Accessible Rich Internet Applications) tools and protocols to make it possible. It is a novel approach, because it is not common that web interfaces allows users to make decisions and to configure the display during the interaction. Concerning that, collapsible panel were used, maintaining the principal framework with few functionalities and giving the user the power to decide what they want to display and use. 

In consideration of how people with visual impairments handling the digital interface, and how screen reader read the graphic user interface, sonoUno web design use the ARIA standard. Not only the ARIA-labels were indicated, but also an specific order to generate a good workflow thought functionalities and ensuring that the screen reader describe the things just as the visual display indicates. Moreover, the unnecessary elements of the visual display are not read by the screen reader, for example, the plot are not read as plot, instead the button play allows to sonify the data plotted.

Another big challenge for this development was to ensure the synchronization between the audio and visual graph, bearing in mind the asynchronous nature of JavaScript. Events with timer were used to guarantee the correct relationship during the reproduction of the data set. Furthermore, during the last tests using large data sets a new problem arise, in the web version and with all this functionalities to plot and sonify large data sets is very difficult, take a lot of time and produce errors in some cases. To solve this issue a decimating filter is being tested.

\subsection{Graphic User Interface design}

In order to maintain the web display as similar as possible to the desktop deployment, a menu was constructed at the top containing: input (allows to open csv/txt data sets, sound and marks that could be done in a data set pointing to parts of interest in the data); output (allows to save the sound, png plot and marks); sample data (this menu item contain pre-loaded data sets that can be displayed on the tool); help (open the complete manual of the tool); and quickstart (open a summary of what to expect and how to use the principal functions).

The reproduction buttons are always displayed and under the plot, these buttons are: play/pause, stop, mark point, delete mark, reset plot, the command text entry and the two sliders to indicate the x position and the tempo. On the other hand, math functionalities and the configurations are located on collapsible panels, this allows to maintain an organized display and few elements that have to be read by the screen reader (helps to reduce memory overload). 

The sound and graphic display can be customized by the end user as their desire. About the sound the maximum and minimum frequency can be set, the volume, the sound type (sine, flute, piano and celesta), choose between continuous and logarithmic scale, and the envelope of the sound. Secondly, the plot configuration allows to set the titles, the grid, the line, the markers, and to flip the x and y axis.

\section{Results}

\begin{figure}
\includegraphics[width=\textwidth]{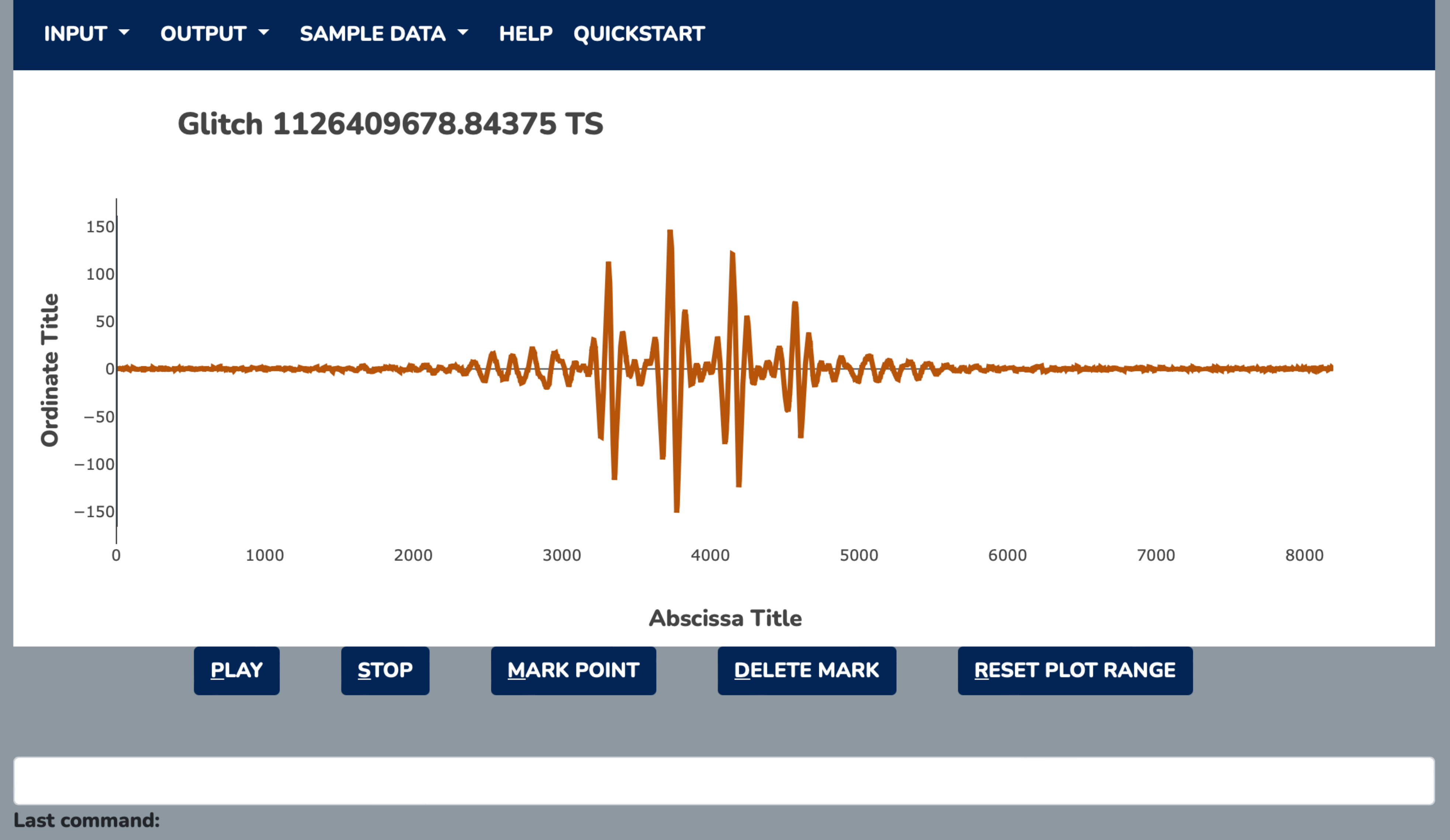}
\caption{A sonoUno web interface screenshot, it include the menu, plot, reproduction buttons and the command line text box. The plot shows a gravitational wave glitch, detected by EGO\cite{ego-zooniverse} and part of the open data provided by the REINFORCE project} \label{fig1}
\end{figure}

A screenshot of the interface was shown in Figure \ref{fig1}. This web tool allows users to see and hear data sets opened from csv or txt files, also end users can load data sets from 'Data Sample' menu item, for example the gravitational wave glitch showed in Figure \ref{fig1} was selected from that menu item. At the bottom, the text entry box, allows to write the functionalities available on the interface (this feature allows to use the web interface from there avoiding the use of the mouse).

\begin{figure}
    \includegraphics[width=\textwidth]{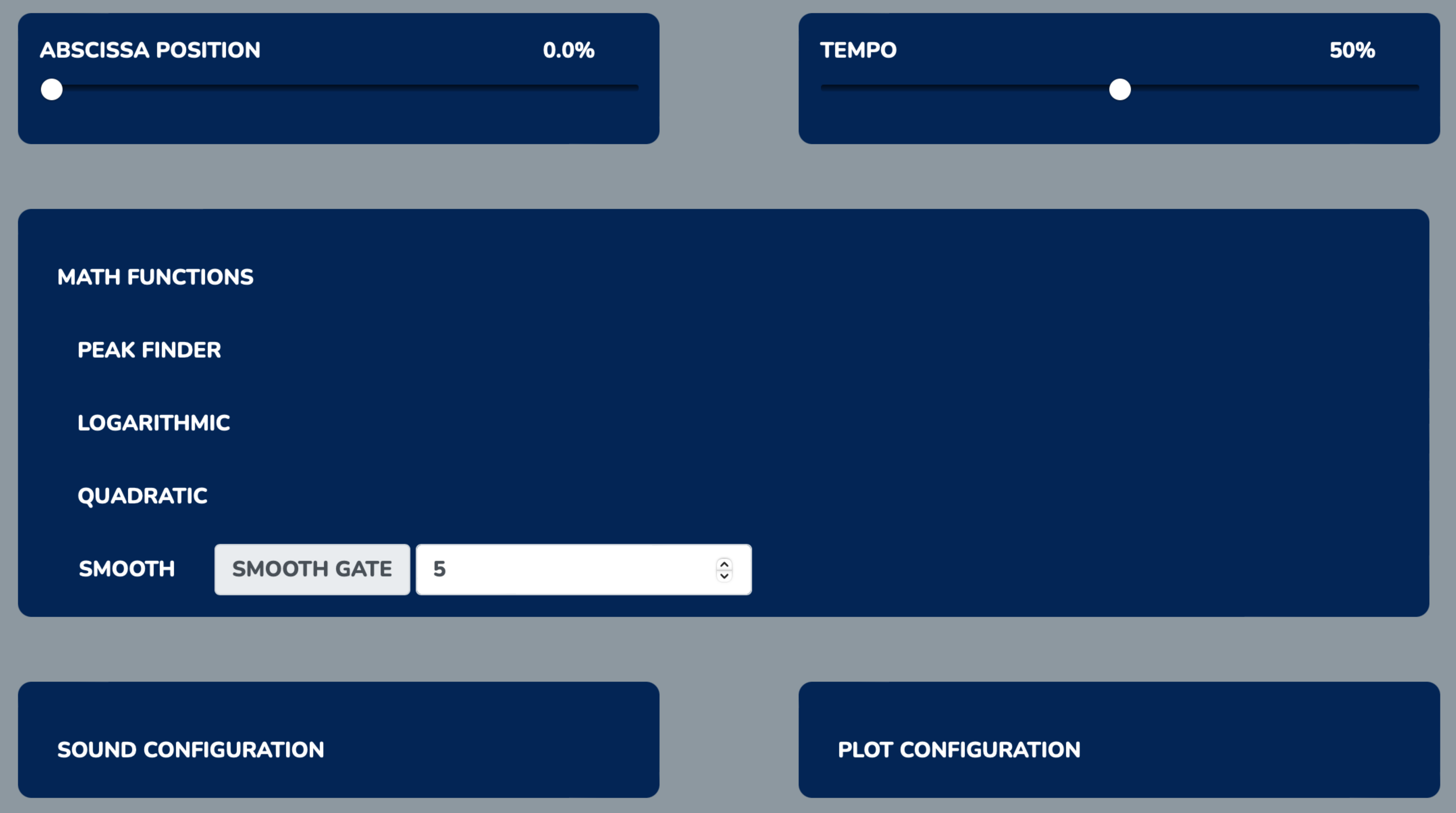}
    \caption{A screenshot with the x position and tempo sliders at the top, the math function panel opened and the sound and plot configurations collapsed.}
    \label{fig2}
\end{figure}

\begin{figure}
    \includegraphics[width=\textwidth]{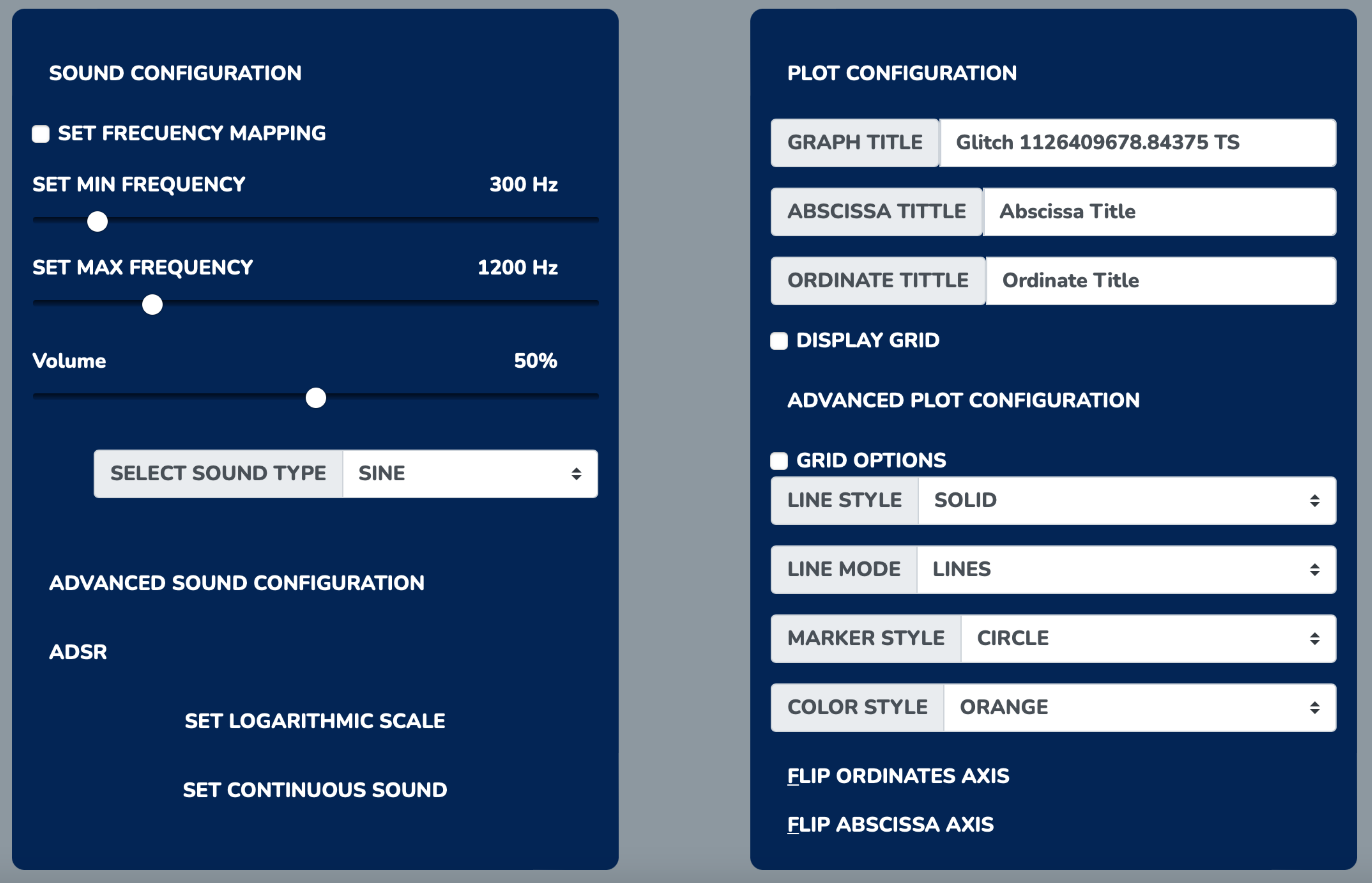}
    \caption{A screenshot of sound and plot configurations panels opened.}
    \label{fig3}
\end{figure}

The plot section allows to zoom it directly on the same plot with the mouse. The abscissa position slider (see Figure \ref{fig2} at the top) allows to move the cursor through the data set and to begin the reproduction from there. The tempo slider allows to speed up and down the reproduction time. Figure \ref{fig2} also shows opened the math function panel, where at the moment there are four functions ready to use: peak finder (in this case a new window allows to select the percentage of sensitivity and if you want to clean or mark the peaks); logarithmic; quadratic; and smooth. At bottom of Figure \ref{fig2} the configuration panels collapsed are shown. Figure \ref{fig3} shows the configuration panels opened with all it functions.

The SonoUno web interface was tested in different platforms with different screen readers (NVDA on Windows, Voice Over on MAC and Orca on Ubuntu). All the elements are enunciated by the screen reader, the elements on the panels are only recognizable when the panel is opened (this is very important to maintain the relation between the visual and auditory display).

\section{Conclusion}

A web interface of sonoUno software was developed, maintaining the original functionalities distribution as same as possible, continuing the user center design from the beginning of the tool. This web interface allows the user to explore the data set, making decisions about what the want to display and how.

Concerning the use of screen readers, the elements of the interface present descriptions and the order of the audible display was carefully design, to ensure the adequate correlation between visual and audible deployment. The principal free screen reader of each operative system was tested, and the results show a good performance.

This innovative approach seeks to continue growing, removing barriers and offering more accessible tools to analyse data sets. Since the beginning of the year, this web interface is being used by a professor at Spain with visual impaired students. This experience will let us know some features to enhance and if the sonoUno web interface can be use by student to better understand math and science. 

As future works, the web interface is adapting to be used from any mobile device, and the axis limits of the plot will be set indicating the specific number to cut. New user tests and focus group will be performed to maintain and assure the user centred design philosophy of sonoUno and all the associated tools. 

%
%
%
%

\end{document}